\documentclass[prb,twocolumn,showpacs,superscriptaddress,preprintnumbers,amssymb]{revtex4}
\usepackage{graphicx}
\usepackage{dcolumn}
\usepackage{bm}
\usepackage{wasysym}

\newcommand{\beq}{\begin{equation}}
\newcommand{\eeq}{\end{equation}}
\newcommand{\beqn}{\begin{eqnarray}}
\newcommand{\eeqn}{\end{eqnarray}}

\begin{document}
\title{Quantum Phase Transitions beyond the Landau's Paradigm in Sp(4) Spin System}
\author{Yang Qi and Cenke Xu}
\affiliation{Department of Physics, Harvard University, Cambridge,
MA 02138}
\date{\today}

\begin{abstract}

We propose quantum phase transitions beyond the Landau's paradigm
of Sp(4) spin Heisenberg models on the triangular and square
lattices, motivated by the exact Sp(4)$\simeq$ SO(5) symmetry of
spin-3/2 fermionic cold atomic system with only $s-$wave
scattering. On the triangular lattice, we study a phase transition
between the $\sqrt{3}\times\sqrt{3}$ spin ordered phase and a
$Z_2$ spin liquid phase, this phase transition is described by an
O(8) sigma model in terms of fractionalized spinon fields, with
significant anomalous scaling dimensions of spin order parameters.
On the square lattice, we propose a deconfined critical point
between the Neel order and the VBS order, which is described by
the CP(3) model, and the monopole effect of the compact U(1) gauge
field is expected to be suppressed at the critical point.

\end{abstract}
\pacs{} \maketitle

\section{Introduction}

Landau's classic phase transition paradigm describes continuous
phase transitions by symmetry breaking of the system
\cite{landau}, and the powerful renormalization group theory
developed by Wilson suffices this paradigm with systematic
calculation techniques. Based on Landau-Ginzburg-Wilson (LGW)
theory \cite{wilson1974}, the continuous phase transition should
be described by fluctuations of physical order parameters. A few
years ago, it was proposed that a direct unfine-tuned continuous
transition between two ordered phases which break different
symmetries is possible in quantum magnet
\cite{senthil2004,senthil2004a}, which is forbidden in Landau's
theory. Recent numerical results suggest that this transition may
exist in a SU(2) spin-1/2 model with both Heisenberg and ring
exchange \cite{sandvik2007,kaul2007}. The key feature of this
nonlandau critical behavior is that at the critical point the
field theory in terms of fractionalized objects with no obvious
physical probe is a more appropriate description. In spite of the
difficulty of probing the fractionalized excitations, the
fractionalized nature of the critical point leads to enormous
anomalous dimension of the physical order parameter distinct from
the Wilson-Fisher fixed point or the mean field result, which can
be checked experimentally.

In a seminal paper, it was proved that in spin-3/2 cold atom
systems, with the standard $s-$wave scattering approximation, the
four-component spin-3/2 fermion multiplet enjoys an enlarged
Sp(4)$\simeq$ SO(5) symmetry without fine-tuning any parameter
\cite{wu2003}. By tuning the spin-0 and spin-2 scattering
channels, there is one point with an even larger SU(4)$\supset$
Sp(4) symmetry \cite{wu2003,wu2005a,wu2006b}. The fundamental
representation of the 15 generators of SU(4) Lie-algebra can be
divided into two groups: $\Gamma_a$ with $a = 1,2 \cdots 5$ and
$\Gamma_{ab} = \frac{1}{2i}[\Gamma^a, \Gamma^b]$, and $\Gamma_a$
obey the Clifford algebra: $\{ \Gamma^a, \Gamma^b \} =
2\delta_{ab}$. Let us denote the fermion atom operator as
$\psi_\alpha$, then the fermion bilinear $\hat{\Gamma}_a =
\psi^\dagger \Gamma_a \psi$ form a vector representation of Sp(4)
group, and $\hat{\Gamma}_{ab} = \psi^\dagger \Gamma_{ab} \psi$
form an adjoint representation of Sp(4) group. In the particular
representation we choose, \beqn \Gamma_a = \sigma^a\otimes \mu^z,
\ a = 1,2,3, \ \Gamma^4 = 1\otimes \mu^x, \Gamma^5 = 1\otimes
\mu^y. \eeqn The difference between SU(4) algebra and Sp(4)
algebra is that, two Sp(4) particles can form a Sp(4) singlet
through a $4\times 4$ antisymmetric matrix $\mathcal{J} =
i\sigma^y\otimes \mu^x$, which satisfies the following algebra
\beqn \mathcal{J}^t = - \mathcal{J}, \ \mathcal{J}^2 = -1, \
\mathcal{J}\Gamma_{ab}\mathcal{J} = \Gamma_{ab}^t, \
\mathcal{J}\Gamma_{a}\mathcal{J} = - \Gamma_{a}^t.
\label{identity}\eeqn One can see that
$\mathcal{J}_{\alpha\beta}\psi^\dagger_{\alpha}\psi^\dagger_{\beta}$
creates a Sp(4) invariant state, therefore the Valence Bond Solid
(VBS) state of SU(2) spin systems can be naturally generalized to
Sp(4) spin systems. By contrast, two SU(4) particles can only form
a 6 dimensional representation and a 10 dimensional representation
of SU(4) algebra, and the smallest SU(4) singlet always involves
four particles.

If we consider a Mott-Insulator phase of spin-3/2 cold atoms on
the optical lattice with one particle per well on average, the
effective spin Hamiltonian should be invariant under Sp(4)
transformations. The most general Sp(4)-Heisenberg model contains
two terms: \beqn H =
\sum_{<i,j>}J_1\hat{\Gamma}^{ab}_i\hat{\Gamma}^{ab}_j - J_2
\hat{\Gamma}^a_i\hat{\Gamma}^a_j. \label{heisenberg}\eeqn The key
difference between $\hat{\Gamma}_{ab}$ and $\hat{\Gamma}_a$ is
their behavior under time-reversal transformation. The
time-reversal transformation on the fermion multiplet
$\psi_\alpha$ is $ \psi_\alpha \rightarrow
\mathcal{J}_{\alpha\beta}\psi_\beta$, this implies that
$\hat{\Gamma}_{ab}$ ($\hat{\Gamma}_a$) is odd (even) under
time-reversal. Also, if rewritten in terms of the original SU(2)
spin-3/2 matrices, $\Gamma_{ab}$ only involves the odd powers of
spins, and $\Gamma_a$ only involves the even powers of spins
\cite{wu2006b}. This model can be exactly realized in spin-3/2
cold atom systems, the coefficients $J_1$ and $J_2$ are determined
by the spin-0 and spin-2 scattering parameters  \cite{wu2006b}.
Clearly when $ - J_2 = J_1$ the system has SU(4) symmetry. In this
work we will consider the Heisenberg model on the triangular and
square lattice, in the parameter regime with $J_1 > 0$. Our focus
in the current work will be the nonlandau like quantum phase
transitions, which is also a larger spin generalization of the
deconfined criticality discussed before. A more detailed analysis
of the whole phase diagram of the Sp(4) Heisenberg model in
(\ref{heisenberg}) will be given in a future work \cite{qifuture}.

\section{the Sp(4) Heisenberg model on the triangular lattice}

\begin{figure}
\includegraphics[width=3.0in]{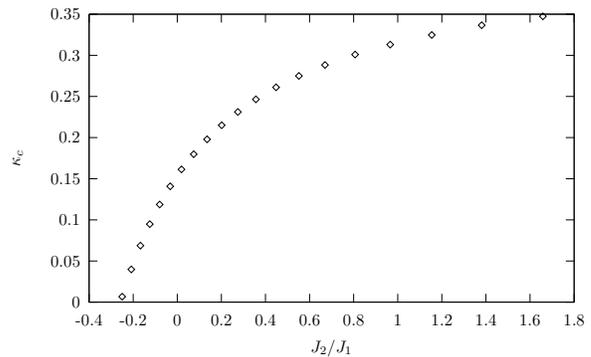}
\caption{Mean field solutions on triangular lattice at different
$J_2/J_1$ with $J_1>0$. The $y$ axes $\kappa_c$ shows the density
of spinon condensate, which is also proportional to $|z|^2$
defined in Eq. \ref{slowmode}. $\kappa_c$ decreases to nearly zero
(0.007) as $J_2/J_1$ decreases.} \label{fig:mft}\end{figure}

Let us study the triangular lattice first, and we will use the
standard Schwinger boson formalism to study the magnetic ordered
phase. We introduce Schwinger boson spinon $b_\alpha$ as usual
$\hat{S}^{a}_i = b^\dagger_{i,\alpha} S^a_{\alpha\beta}b_{i,\beta}
$, $\hat{S}^a$ are the 15 generators of SU(4) algebra in the
fundamental representation. This definition of spinon $b_\alpha$
is subject to a local constraint: $\sum_{\alpha = 1}^4
b^\dagger_{i,\alpha}b_{i,\alpha} = 1$, which also manifests itself
as a local U(1) degree of freedom: $b_{i,\alpha}\rightarrow
\exp(i\theta_i)b_{i,\alpha}$. Using the following identities
\cite{xu2008}: \beqn
\Gamma^{ab}_{\alpha\beta}\Gamma^{ab}_{\gamma\sigma} &=& 2
\delta_{\alpha\sigma}\delta_{\beta\gamma} - 2
\mathcal{J}_{\alpha\gamma}\mathcal{J}_{\beta\sigma}, \cr \cr
\Gamma^{a}_{\alpha\beta}\Gamma^{a}_{\gamma\sigma} &=& 2
\delta_{\alpha\sigma}\delta_{\beta\gamma} + 2
\mathcal{J}_{\alpha\gamma}\mathcal{J}_{\beta\sigma} -
\delta_{\alpha\beta}\delta_{\gamma\sigma}, \eeqn the Hamiltonian
(\ref{heisenberg}) can be rewritten as \beqn H = \sum_{<i,j>}2(J_1
- J_2)\hat{K}^\dagger_{ij}\hat{K}_{ij} -2(J_1 +
J_2)\hat{Q}_{ij}^\dagger \hat{Q}_{ij}, \cr\cr \hat{K}_{ij} =
b^\dagger_{i,\alpha}b_{j,\alpha}, \ \ \hat{Q}_{ij} =
\mathcal{J}_{\alpha\beta}b_{i,\alpha}b_{j,\beta}. \eeqn Now we
introduce two variational parameters $K_{ij} = \langle
\hat{K}_{ij} \rangle$ and $Q_{ij} = \langle \hat{Q}_{ij} \rangle$,
and assuming these variational parameters are uniform on the whole
lattice, we meanfield Hamiltonian for (\ref{heisenberg}) reads:
\beqn H_{mf} = \sum_{<i,j>} 2(J_1 - J_2)K \hat{K}_{ij} - 2(J_1 +
J_2) Q \hat{Q}_{ij} + H.c. \cr\cr -2(J_1 - J_2)K^2 + 2(J_1 +
J_2)Q^2 - \mu(b^\dagger_{i,\alpha}b_{i,\alpha} - 1). \eeqn The
following formalism is similar to reference \cite{wang2007}, which
studied the SU(2) spin models on the triangular lattice. The term
involving $\mu$ imposes the constraint on the Hilbert space of
spinon $\sum_{\alpha = 1}^4 b^\dagger_{i,\alpha}b_{i,\alpha} = 1$.
If the spectrum of the spinons is gapless, the spinon will
condense at the minima of the Brillouin zone. By solving the
self-consistent equations for $K$, $Q$ and $\mu$, we obtain that
when $J_2/J_1 > - 0.3$, there is a finite percentage of spinon
condensate at momenta $\pm \vec{q}_0 = \pm (\frac{2\pi}{3},
\frac{2\pi}{\sqrt{3}})$, which are the corners of the Brillouin
Zone. The condensate density as a function of $J_2/J_1$ is plotted
in Fig. \ref{fig:mft}.

The gauge field fluctuation rooted in the constraint $\sum_{\alpha
= 1}^4 b^\dagger_{i,\alpha}b_{i,\alpha} = 1$ is the most important
correction to the mean field calculation above. The local
constraint would in general induce U(1) gauge fluctuations.
However, the condensate obtained from the Schwinger boson
formalism corresponds to the state with nonzero expectation value
$Q = \langle \hat{Q}_{ij}\rangle$, which is a pairing amplitude.
The pairing between nearest neighbor sites breaks the U(1) gauge
symmetry down to $Z_2$ gauge symmetry, therefore the long
wavelength field theory of this condensate should only have $Z_2$
gauge symmetry. To understand this order, we define slow mode
$z_\alpha$ as \beqn b_\alpha(x) = e^{i\vec{q}_0\cdot \vec{x}}
z_\alpha(x) + e^{- i\vec{q}_0\cdot \vec{x}}
\mathcal{J}_{\alpha\beta}z^\ast_\beta(x),\label{slowmode} \eeqn
now one can rewrite spin operators $\hat{\Gamma}_{ab}$ and
$\hat{\Gamma}_a$ in terms of slow mode $z_\alpha$ as \beqn
\hat{\Gamma}_{ab} & \sim & e^{i 2\vec{q}_0\cdot\vec{x}}
z\mathcal{J}\Gamma_{ab}z
 + H.c., \cr \cr \hat{\Gamma}_{a} &\sim &
z^\dagger\Gamma_a z = n_a. \eeqn Therefore $\hat{\Gamma}_a$ has a
uniform order $n_a$, while $\hat{\Gamma}_{ab}$ is only ordered at
finite momentum $\pm 2\vec{q}_{0}$. For completeness, one can
define Sp(4) adjoint vector $n_{1,ab}$ and $n_{2,ab}$ as \beqn
n_{1,ab} &=& \mathrm{Re}[z\mathcal{J}\Gamma_{ab}z], \ n_{2,ab} =
\mathrm{Im}[z\mathcal{J}\Gamma_{ab}z],  \eeqn the order of
$\hat{\Gamma}_{ab}$ can be written in terms of $n_{1,ab}$ and
$n_{2,ab}$: \beqn \hat{\Gamma}_{ab} \sim
\cos(2\vec{q}_0\cdot\vec{x})n_{1,ab} +
\sin(2\vec{q}_0\cdot\vec{x})n_{2,ab}, \cr \cr
\sum_{a,b}n_{1,ab}n_{2,ab} = 0. \label{sincos}\eeqn $n_{1,ab}$ and
$n_{2,ab}$ are two Sp(4) adjoint vectors ``perpendicular" to each
other. Since $\hat{\Gamma}_a$ is time-reversal even, while
$\hat{\Gamma}_{ab}$ is time-reversal odd \cite{wu2006b}, the
condensate of $z_\alpha$ has both uniform spin nematic order and
$\sqrt{3}\times \sqrt{3}$ order.

The U(1) local gauge degree of freedom is lost in Eq.
\ref{slowmode}, The residual gauge symmetry is only $Z_2$ which
transforms $z\rightarrow -z$. Physically this implies that an
arbitrary U(1) transformation of $z$ field will result in a
rotation of spin order parameter $\hat{\Gamma}_{ab}$. This
situation is very similar to the spinon description of the
$\sqrt{3}\times \sqrt{3}$ order of SU(2) spins on the triangular
lattice \cite{senthil1994}. The field theory describing this
condensate should contain $Z_2$ gauge field. However, since $Z_2$
gauge field does not introduce any long range interaction or
critical behavior, we can safely integrate out the $Z_2$ gauge
field. The field theory can then be written as \beqn L =
|\partial_\mu z|^2 + r|z|^2 + g(|z|^2)^2 + \cdots
\label{action}\eeqn The ellipses include all the Sp(4) invariant
terms.

Apparently, without the ellipses, the Lagrangian (\ref{action})
enjoys an enlarged O(8) symmetry once we define real boson field
multiplet $\vec{\phi}$ as $\vec{\phi} = (\mathrm{Re}[z_1],
\mathrm{Im}[z_1], \cdots, \mathrm{Im}[z_4])^t$, and the Lagrangian
(\ref{action}) can be rewritten as \beqn L = \sum_{\alpha =
1}^8(\partial_\mu\phi_\alpha)^2 + r |\vec{\phi}|^2 +
g(|\vec{\phi}|^2)^2 + \cdots \label{action1} \eeqn The Lagrangian
(\ref{action1}) without other perturbations describes an O(8)
transition, and the ordered state has ground state manifold (GSM)
\beqn \mathrm{U(4)}/[\mathrm{U(3)}\otimes Z_2] = S^7/Z_2 =
\mathrm{RP(7)}, \eeqn we mod $Z_2$ from $S^7$ because of the $Z_2$
gauge symmetry of $z$. There are certainly other terms in the
field theory which can break the O(8) symmetry down to Sp(4)
symmetry, but all the terms allowed by Sp(4) symmetry and lattice
symmetry include at least two derivatives, for instance $
|\mathcal{J}_{\alpha\beta}z_\alpha\partial_\mu z_\beta|^2 $. These
terms change the Goldstone mode dispersion but do not change the
GSM, and since they contain high powers of $z$ and also at least
two derivatives, they are irrelevant at the O(8) critical point.
Other Sp(4) invariant terms without derivatives like
$\sum_{a,b}(n_{1,ab})^2$, $\sum_{a}(n_{a})^2$,
$\epsilon_{abcde}n_{1,ab}n_{1,cd}n_e$ \emph{et al.} either vanish
or can be rewritten in terms of powers of $z^\dagger z$, which
preserves the O(8) symmetry. Therefore we conclude that the ground
state manifold of the condensate is $S^7/Z_2$, and the transition
between the condensate and disordered state by tuning $J_1/J_2$
belongs to the O(8) universality class. This transition is beyond
the Landau's paradigm in the sense that the field theories
(\ref{action}) and (\ref{action1}) are written in terms of spinon
field instead of physical order parameters. The physical order
parameters are bilinears of spinon, which implies that the
anomalous dimension of the physical order parameters are enormous
at this transition.

Since the GSM is $S^7/Z_2$ with fundamental group $\pi_1[S^7/Z_2]
= Z_2$, in the condensate there are gapped visons, which is a
``$\pi-$flux" of the ``Higgsed" $Z_2$ gauge field. The disordered
phase is actually a $Z_2$ spin liquid with gapped but mobile
visons. This $Z_2$ spin liquid phase of SU(2) spin systems can be
most conveniently visualized in the quantum dimer model (QDM) on
the triangular lattice \cite{sondhi2001}, which by tuning the
dimer flipping and dimer potential energy, stabilizes a gapped
phase with $Z_2$ topological order and no symmetry breaking. As we
discussed earlier, two Sp(4) particles can form a Sp(4) singlet,
therefore the QDM for Sp(4) spin systems is exactly the same as
the SU(2) spins, with also a stable $Z_2$ spin liquid phase.
Because the $Z_2$ spin liquid is a deconfined phase, the
excitations of the $Z_2$ spin liquid include gapped Sp(4) bosonic
spinons besides the visons, If we start with the disordered $Z_2$
spin liquid state, and drive a transition by condensing the gapped
Sp(4) spinon, the field theory of this transition is in the same
form as (\ref{action}).

Since on one site there is only one particle, the particular QDM
is subject to the local constraint with one dimer connected to
each site. This type of QDM is called odd QDM, since the product
of $Z_2$ electric field around each site is $\prod \sigma^x = -1$,
which will attach a $\pi-$flux to each hexagon of the dual
honeycomb lattice of the triangular lattice. This $\pi-$flux seen
by the visons will lead to four degenerate minima in the vison
band, and the condensation of the vison at these minima breaks the
translation and rotation symmetry of the lattice
\cite{sondhi2001a}, and the transition has been suggested to be an
O(4) transition.

In this section we discussed the transition between the $Z_2$ spin
liquid and the $\sqrt{3}\times \sqrt{3}$ state of the Sp(4) spin
system. For comparison, let us briefly discuss the order-disorder
transition of the $\sqrt{3}\times \sqrt{3}$ state in the standard
Landau theory, ignoring the topological nature of the $Z_2$ spin
liquid. In the $\sqrt{3}\times \sqrt{3}$ order, both time-reversal
and Sp(4) spin symmetries are broken. A general Ginzburg-Landau
Lagrangian can be written in terms of the time-reversal even O(5)
vector $n^a$ which is defined as the long wavelength field of
$\hat{\Gamma}^a$, and two adjoint vectors $n^{ab}_{1}$ and
$n^{ab}_{2}$ introduced in Eq. \ref{sincos}. At the quadratic
level none of these three vectors mix, while at the cubic order a
mixing term is allowed by the Sp(4) symmetry: $ \sum_{i = 1}^2
\epsilon_{abcde}n^{ab}_{i}n^{cd}_i n^e $, this term implies that
the ordering of the adjoint vectors would drive the order of
$n^a$, but the statement is not necessarily true conversely. If
the O(5) vector $n_a$ is ordered while the adjoint vectors
$n^{ab}_1$, $n^{ab}_2$ are disordered, the system breaks the Sp(4)
symmetry while preserving the time-reversal symmetry. Therefore if
the system is tuned towards the disordered phase, the Landau's
theory allows for multiple transitions, with the time-reversal
symmetry restored before the Sp(4) symmetry. A uniform collinear
order $\hat{\Gamma}_a$ has GSM $ S^4 =
\mathrm{SO(5)}/\mathrm{SO(4)}$, therefore the transition of $n^a$
belongs to the O(5) universality class. The transition associated
with time-reversal symmetry breaking is described by the O(10)
vectors $n^{ab}_1$ and $n^{ab}_2$, with various anisotropies in
the background of the gapless O(5) ordering $n^a$. For instance,
at the quartic order, there is a term which imposes the
``orthogonality" between the two O(10) vectors: $(\sum_{a,b}
n_1^{ab}n_2^{ab})^2$. The nature of this transition requires more
detailed analysis. By contrast, the Schwinger boson and field
theory analysis show that there can be a direct O(8) transition
between the phase with coexistence of $n_a$ and $n_{i,ab}$, and a
spin disordered phase with $Z_2$ topological order.

\section{Sp(4) Heisenberg model on the square lattice}

Now let us switch the gear to the square lattice. On the square
lattice, at the point with $J_1 = J_2 > 0$, the model
(\ref{heisenberg}) can be mapped to the SU(4) Heisenberg model
with fundamental representation on one sublattice and conjugate
representation on the other \cite{wu2006b}. The equivalence can be
shown by performing transformation $S^a\rightarrow
\mathcal{J}^\dagger S^a \mathcal{J}$ on one of the sublattices,
and using the identities in (\ref{identity}). This point $J_1 =
J_2$ has been thoroughly studied by means of large-$N$
generalization \cite{sachdev1990,sachdev1989,arovas1988} and
quantum Monte Carlo \cite{kawashima2003}. It is agreed that at
this point the spinon $b_\alpha$ condenses, and there is a small
Neel moment \cite{gmzhang2001,kawashima2003}. In the Schwinger
boson language, the Neel state on the square lattice corresponds
to the condensate of Schwinger bosons with nonzero expectation of
$\langle\hat{Q}_{ij}\rangle$, which seems to break the U(1) gauge
symmetry down to $Z_2$. However, the U(1) gauge symmetry can be
restored if the Schwinger bosons on the two sublattices are
associated with opposite gauge charges, therefore the connection
between spinon $b_\alpha$ and low energy field $z_\alpha$ is \beqn
b_{\alpha} &\sim& z_\alpha, \ (\mathrm{sublattice \ A}), \cr\cr \
b_{\alpha} &\sim& \mathcal{J}_{\alpha\beta}z^\dagger_\beta, \
(\mathrm{sublattice \ B}). \eeqn The GSM of the Schwinger boson
condensate is \beqn \mathrm{U(4)}/[\mathrm{U(1)}\otimes
\mathrm{U(3)}] = S^7 / \mathrm{U(1)} = \mathrm{CP(3)}.\eeqn The
field theory for this condensate is most appropriately described
by the CP(3) model \beqn L = \sum_{\alpha = 1}^4 |(\partial_\mu -
ia_\mu)z_\alpha|^2 + r|z|^2 + g(|z|^2)^2 + \cdots \label{cp3}\eeqn
Again, if we perturb this field theory with Sp(4) invariant terms,
the GSM is still CP(3), and the critical behavior is unchanged.
The condensate of $z_\alpha$ has staggered spin order
$\hat{\Gamma}_{ab}$ but uniform nematic order $\hat{\Gamma}_{a}$
on the square lattice.

In the condensate of $z$, gauge field $a_\mu$ is Higgsed; if $z$
is disordered, $a_\mu$ would be in a gapless photon phase if the
gauge fluxes are conserved. However, because $
\pi_2[\mathrm{CP}(3)] = Z$, the ground state manifold can have
singular objects in the 2+1 dimensional space time
\cite{sachdev1990}, which corresponds to the monopole of the
compact U(1) gauge field $a_\mu$. The conservation of gauge fluxes
is broken by the monopoles, which due to its Berry phase will
drive the system to a phase breaking the lattice symmetry
\cite{haldane1988a,sachdev1990}.

\begin{figure}
\includegraphics[width=3.0in]{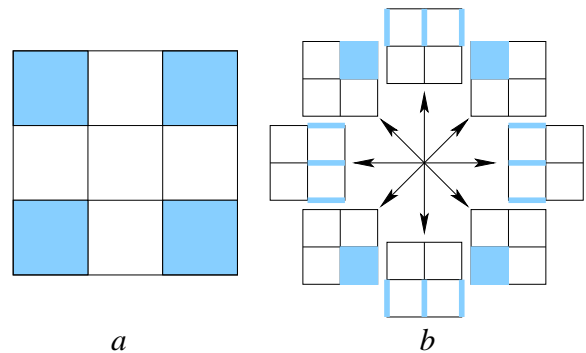}
\caption{Fig. $a$, the SU(4) plaquette order pattern, four SU(4)
particles around the colored squares form a SU(4) singlet; Fig.
$b$ the particular type of VBS state depends on the phase angle of
the monopole operator.} \label{plaquette}
\end{figure}

At another point with $-J_2 = J_1 > 0$, this model is SU(4)
invariant with fundamental representations on both sublattices.
This point is not so well studied. A fermionic mean field theory
\cite{fczhang2002} and an exact diagonalization \cite{fczhang2000}
on a $4\times 4$ lattice has been applied to this point, the
results suggest that the ground state may be a plaquette order as
depicted in Fig. \ref{plaquette}, with four particles forming a
SU(4) singlet on every one out of four unit squares. A similar
plaquette ordered phase is obtained on the spin ladder
\cite{wu2005}. It is interesting to consider the dynamics of the
plaquettes, for instance, in 3 dimensional cubic lattice, a
quantum plaquette model as a generalization of the quantum dimer
model has been studied both numerically \cite{pankov2007} and
analytically \cite{xuwu2008}. If we perturb away from the SU(4)
point with Sp(4) invariant terms, this plaquette order is expected
to persist into a finite region in the phase diagram due to its
gapped nature. This phase presumably can be continuously connected
to the Sp(4) VBS state with Sp(4) singlets resonating on every one
of four unit squares (Fig. \ref{plaquette}), because both states
are gapped and break the same lattice symmetry. More details about
the possible phases on the square lattice is under study by
another group \cite{congjun}.

The dimer resonating plaquette state can be understood in the same
way as the dimer columnar state as the proliferation of monopoles
of the compact U(1) gauge field, and the oscillating Berry phase
of the monopoles will choose the specific lattice symmetry
breaking pattern. Both the dimer columnar order and the dimer
plaquette order can be viewed as a condensate of fluxes of U(1)
gauge field with the U(1) conservation of fluxes breaking down to
$Z_4$, and if the phase angle of the condensate is $2n \pi/4$ the
system is in the columnar state, while if the phase angle is $(2n
+ 1) \pi/4$ the system is in the dimer plaquette phase
\cite{sachdev1990} (Fig. \ref{plaquette}). If one considers a pure
QDM on the square lattice, the crystalline pattern can be obtained
from the dual rotor model with Lagrangian
\cite{fradkin1990,sachdev1991}: \beqn L_{d} = (\partial_\mu
\chi)^2 - \alpha\cos(8\pi\chi). \eeqn Here $\exp(i2\pi \chi)$ is
the monopole operator which creates a $2\pi$ flux of the U(1)
gauge field. Now whether the system favors columnar order or dimer
resonating plaquette order simply depends on the sign of $\alpha$.

\begin{figure}
\includegraphics[width=2.8in]{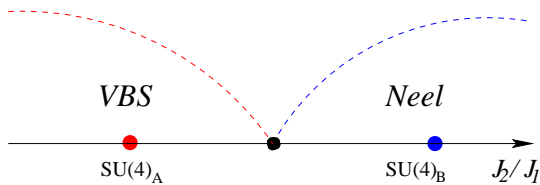}
\caption{The conjectured phase diagram for the Heisenberg model in
equation (\ref{heisenberg}). By tuning one parameter $J_2 / J_1$,
the system evolves from the Neel state to the dimerized VBS state,
and the transition can be continuous. The red and blue dashed line
denotes the magnitude of the Neel and VBS order parameter.
$\mathrm{SU(4)_A}$ is the SU(4) invariant point with fundamental
representation on all sites; $\mathrm{SU(4)_B}$ point is another
SU(4) point with fundamental representation on one sublattice, and
conjugate representation on the other.} \label{phasedia}
\end{figure}

Now we conjecture a phase diagram (Fig. \ref{phasedia}): suppose
$J_1$ is fixed, and we tune $J_2$; if $J_2 > J_c$ the system
remains in the condensate of $z$, which is the Neel order of spin
operators; when $J_2 < J_c$ the system loses the Neel order and
enters the VBS state. This transition can be a direct second order
transition, and the field theory is described by the CP(3) model
in (\ref{cp3}), assuming the CP(3) model itself has a second order
transition. The most important instability on this field theory is
the monopole of the compact U(1) gauge field, which is certainly
relevant in the crystalline phase. However, it has been shown
convincingly that at the 3D XY transition, the $Z_4$ anisotropy of
the XY variable is irrelevant \cite{sandvik2007a}, and it was also
argued that large number of flavor of boson field tends to
suppress the monopole effects \cite{senthil2004a}, therefore it is
likely that the monopoles are irrelevant at the critical point
described by field theory (\ref{cp3}). Compared with the SU(2)
spin system, our Sp(4) system with doubled number of complex boson
fields has a better chance to ensure the irrelevance of monopole
perturbations at the transition. We also want to point out that
between the Neel order and the dimer resonating plaquette order,
an intermediate phase with columnar order is also possible. But
the transition between the Neel order and the columnar order is
also described by field theory (\ref{cp3}), and the columnar order
is connected to the resonating plaquette order through a first
order transition.

\section{summary and extension}

In this work we studied the quantum phase transitions beyond the
Landau's paradigm in the spin-3/2 cold atom systems with emergent
enlarged Sp(4) symmetry. Compared with the $J-Q$ model studied
before \cite{kaul2007,sandvik2007}, the spin model we considered
is very realistic, we propose these results are observable in real
experimental systems in future. It would also been interesting to
study the Heisenberg model in this work through numerical
techniques. A careful numerical study of the classical CP(3) model
without monopoles is also desired, as has been done recently for
the SU(2) invariant CP(1) model \cite{ashvin2008}.

The current work focused on the parameter regime with $J_1 > 0$.
In the regime with $J_1 < 0$, the Schwinger boson formalism would
lead to the ordered state with nonzero expectation value $K =
\langle \hat{K}_{ij} \rangle$, and the Schwinger bosons condense
at momentum $(0,0)$. This state is the ferromagnetic state with
uniform order $n^{ab}$ and $n^a$. The ferromagnetic state and the
Neel state can be connected through a first order transition. More
theoretical tools are desired to determine the other parts of the
phase diagram accurately. We will leave this to the future work
\cite{qifuture}.

\begin{acknowledgments}

We thank Subir Sachdev for helpful discussions. Cenke Xu is
supported by the Milton Funds of Harvard University.

\end{acknowledgments}

\bibliography{sptransition}

\begin{thebibliography}{31}
\expandafter\ifx\csname natexlab\endcsname\relax\def\natexlab#1{#1}\fi
\expandafter\ifx\csname bibnamefont\endcsname\relax
  \def\bibnamefont#1{#1}\fi
\expandafter\ifx\csname bibfnamefont\endcsname\relax
  \def\bibfnamefont#1{#1}\fi
\expandafter\ifx\csname citenamefont\endcsname\relax
  \def\citenamefont#1{#1}\fi
\expandafter\ifx\csname url\endcsname\relax
  \def\url#1{\texttt{#1}}\fi
\expandafter\ifx\csname urlprefix\endcsname\relax\def\urlprefix{URL }\fi
\providecommand{\bibinfo}[2]{#2}
\providecommand{\eprint}[2][]{\url{#2}}

\bibitem[{\citenamefont{Landau et~al.}(1999)\citenamefont{Landau, Lifshitz, and
  Pitaevskii}}]{landau}
\bibinfo{author}{\bibfnamefont{L.~D.} \bibnamefont{Landau}},
  \bibinfo{author}{\bibfnamefont{E.~M.} \bibnamefont{Lifshitz}},
  \bibnamefont{and} \bibinfo{author}{\bibfnamefont{E.~M.}
  \bibnamefont{Pitaevskii}}, pp. \bibinfo{pages}{Statistical Physics
  (Butterworth--Heinemann, New York)} (\bibinfo{year}{1999}).

\bibitem[{\citenamefont{Wilson and Kogut}(1974)}]{wilson1974}
\bibinfo{author}{\bibfnamefont{K.~G.} \bibnamefont{Wilson}} \bibnamefont{and}
  \bibinfo{author}{\bibfnamefont{J.}~\bibnamefont{Kogut}},
  \bibinfo{journal}{Phys. Rev.} \textbf{\bibinfo{volume}{12}},
  \bibinfo{pages}{75} (\bibinfo{year}{1974}).

\bibitem[{\citenamefont{Senthil
  et~al.}(2004{\natexlab{a}})\citenamefont{Senthil, Vishwanath, Balents,
  Sachdev, and Fisher}}]{senthil2004}
\bibinfo{author}{\bibfnamefont{T.}~\bibnamefont{Senthil}},
  \bibinfo{author}{\bibfnamefont{A.}~\bibnamefont{Vishwanath}},
  \bibinfo{author}{\bibfnamefont{L.}~\bibnamefont{Balents}},
  \bibinfo{author}{\bibfnamefont{S.}~\bibnamefont{Sachdev}}, \bibnamefont{and}
  \bibinfo{author}{\bibfnamefont{M.~P.~A.} \bibnamefont{Fisher}},
  \bibinfo{journal}{Science} \textbf{\bibinfo{volume}{303}},
  \bibinfo{pages}{1409} (\bibinfo{year}{2004}{\natexlab{a}}).

\bibitem[{\citenamefont{Senthil
  et~al.}(2004{\natexlab{b}})\citenamefont{Senthil, Balents, Sachdev,
  Vishwanath, and Fisher}}]{senthil2004a}
\bibinfo{author}{\bibfnamefont{T.}~\bibnamefont{Senthil}},
  \bibinfo{author}{\bibfnamefont{L.}~\bibnamefont{Balents}},
  \bibinfo{author}{\bibfnamefont{S.}~\bibnamefont{Sachdev}},
  \bibinfo{author}{\bibfnamefont{A.}~\bibnamefont{Vishwanath}},
  \bibnamefont{and} \bibinfo{author}{\bibfnamefont{M.~P.~A.}
  \bibnamefont{Fisher}}, \bibinfo{journal}{Phys. Rev. B}
  \textbf{\bibinfo{volume}{70}}, \bibinfo{pages}{144407}
  (\bibinfo{year}{2004}{\natexlab{b}}).

\bibitem[{\citenamefont{Sandvik}(2007)}]{sandvik2007}
\bibinfo{author}{\bibfnamefont{A.~W.} \bibnamefont{Sandvik}},
  \bibinfo{journal}{Phys. Rev. Lett} \textbf{\bibinfo{volume}{98}},
  \bibinfo{pages}{227202} (\bibinfo{year}{2007}).

\bibitem[{\citenamefont{Melko and Kaul}(2007)}]{kaul2007}
\bibinfo{author}{\bibfnamefont{R.~G.} \bibnamefont{Melko}} \bibnamefont{and}
  \bibinfo{author}{\bibfnamefont{R.~K.} \bibnamefont{Kaul}},
  \bibinfo{journal}{arXiv:0707.2961}  (\bibinfo{year}{2007}).

\bibitem[{\citenamefont{Wu et~al.}(2003)\citenamefont{Wu, Hu, and
  Zhang}}]{wu2003}
\bibinfo{author}{\bibfnamefont{C.}~\bibnamefont{Wu}},
  \bibinfo{author}{\bibfnamefont{J.~P.} \bibnamefont{Hu}}, \bibnamefont{and}
  \bibinfo{author}{\bibfnamefont{S.~C.} \bibnamefont{Zhang}},
  \bibinfo{journal}{Phys. Rev. Lett} \textbf{\bibinfo{volume}{91}},
  \bibinfo{pages}{186402} (\bibinfo{year}{2003}).

\bibitem[{\citenamefont{Wu}(2005)}]{wu2005a}
\bibinfo{author}{\bibfnamefont{C.}~\bibnamefont{Wu}}, \bibinfo{journal}{Phys.
  Rev. Lett.} \textbf{\bibinfo{volume}{95}}, \bibinfo{pages}{266404}
  (\bibinfo{year}{2005}).

\bibitem[{\citenamefont{Wu}(2006)}]{wu2006b}
\bibinfo{author}{\bibfnamefont{C.}~\bibnamefont{Wu}}, \bibinfo{journal}{Mod.
  Phys. Lett. B} \textbf{\bibinfo{volume}{20}}, \bibinfo{pages}{1707}
  (\bibinfo{year}{2006}).

\bibitem[{\citenamefont{Qi and Xu}()}]{qifuture}
\bibinfo{author}{\bibfnamefont{Y.}~\bibnamefont{Qi}} \bibnamefont{and}
  \bibinfo{author}{\bibfnamefont{C.}~\bibnamefont{Xu}},
  \bibinfo{howpublished}{In progress}.

\bibitem[{\citenamefont{Xu}(2008)}]{xu2008}
\bibinfo{author}{\bibfnamefont{C.}~\bibnamefont{Xu}},
  \bibinfo{journal}{arXiv:0803.0794}  (\bibinfo{year}{2008}).

\bibitem[{\citenamefont{Wang and Vishwanath}(2006)}]{wang2007}
\bibinfo{author}{\bibfnamefont{F.}~\bibnamefont{Wang}} \bibnamefont{and}
  \bibinfo{author}{\bibfnamefont{A.}~\bibnamefont{Vishwanath}},
  \bibinfo{journal}{Phys. Rev. B} \textbf{\bibinfo{volume}{74}},
  \bibinfo{pages}{174423} (\bibinfo{year}{2006}).

\bibitem[{\citenamefont{Chubukov et~al.}(1994)\citenamefont{Chubukov, Sachdev,
  and Senthil}}]{senthil1994}
\bibinfo{author}{\bibfnamefont{A.~V.} \bibnamefont{Chubukov}},
  \bibinfo{author}{\bibfnamefont{S.}~\bibnamefont{Sachdev}}, \bibnamefont{and}
  \bibinfo{author}{\bibfnamefont{T.}~\bibnamefont{Senthil}},
  \bibinfo{journal}{Nucl. Phys. B} \textbf{\bibinfo{volume}{426}},
  \bibinfo{pages}{601} (\bibinfo{year}{1994}).

\bibitem[{\citenamefont{Moessner and Sondhi}(2001{\natexlab{a}})}]{sondhi2001}
\bibinfo{author}{\bibfnamefont{R.}~\bibnamefont{Moessner}} \bibnamefont{and}
  \bibinfo{author}{\bibfnamefont{S.~L.} \bibnamefont{Sondhi}},
  \bibinfo{journal}{Phys. Rev. Lett} \textbf{\bibinfo{volume}{86}},
  \bibinfo{pages}{1881} (\bibinfo{year}{2001}{\natexlab{a}}).

\bibitem[{\citenamefont{Moessner and Sondhi}(2001{\natexlab{b}})}]{sondhi2001a}
\bibinfo{author}{\bibfnamefont{R.}~\bibnamefont{Moessner}} \bibnamefont{and}
  \bibinfo{author}{\bibfnamefont{S.~L.} \bibnamefont{Sondhi}},
  \bibinfo{journal}{Phys. Rev. B} \textbf{\bibinfo{volume}{63}},
  \bibinfo{pages}{224401} (\bibinfo{year}{2001}{\natexlab{b}}).

\bibitem[{\citenamefont{Read and Sachdev}(1990)}]{sachdev1990}
\bibinfo{author}{\bibfnamefont{N.}~\bibnamefont{Read}} \bibnamefont{and}
  \bibinfo{author}{\bibfnamefont{S.}~\bibnamefont{Sachdev}},
  \bibinfo{journal}{Phys. Rev. B} \textbf{\bibinfo{volume}{42}},
  \bibinfo{pages}{4568} (\bibinfo{year}{1990}).

\bibitem[{\citenamefont{Read and Sachdev}(1989)}]{sachdev1989}
\bibinfo{author}{\bibfnamefont{N.}~\bibnamefont{Read}} \bibnamefont{and}
  \bibinfo{author}{\bibfnamefont{S.}~\bibnamefont{Sachdev}},
  \bibinfo{journal}{Nucl. Phys. B} \textbf{\bibinfo{volume}{316}},
  \bibinfo{pages}{609} (\bibinfo{year}{1989}).

\bibitem[{\citenamefont{Arovas and Auerbach}(1988)}]{arovas1988}
\bibinfo{author}{\bibfnamefont{D.~P.} \bibnamefont{Arovas}} \bibnamefont{and}
  \bibinfo{author}{\bibfnamefont{A.}~\bibnamefont{Auerbach}},
  \bibinfo{journal}{Phys. Rev. B} \textbf{\bibinfo{volume}{38}},
  \bibinfo{pages}{316} (\bibinfo{year}{1988}).

\bibitem[{\citenamefont{Harada et~al.}(2003)\citenamefont{Harada, Kawashima,
  and Troyer}}]{kawashima2003}
\bibinfo{author}{\bibfnamefont{K.}~\bibnamefont{Harada}},
  \bibinfo{author}{\bibfnamefont{N.}~\bibnamefont{Kawashima}},
  \bibnamefont{and} \bibinfo{author}{\bibfnamefont{M.}~\bibnamefont{Troyer}},
  \bibinfo{journal}{Phys. Rev. Lett} \textbf{\bibinfo{volume}{90}},
  \bibinfo{pages}{117203} (\bibinfo{year}{2003}).

\bibitem[{\citenamefont{Zhang and Shen}(2001)}]{gmzhang2001}
\bibinfo{author}{\bibfnamefont{G.-M.} \bibnamefont{Zhang}} \bibnamefont{and}
  \bibinfo{author}{\bibfnamefont{S.-Q.} \bibnamefont{Shen}},
  \bibinfo{journal}{Phys. Rev. Lett} \textbf{\bibinfo{volume}{87}},
  \bibinfo{pages}{157201} (\bibinfo{year}{2001}).

\bibitem[{\citenamefont{Haldane}(1988)}]{haldane1988a}
\bibinfo{author}{\bibfnamefont{F.~D.} \bibnamefont{Haldane}},
  \bibinfo{journal}{Phys. Rev. Lett} \textbf{\bibinfo{volume}{61}},
  \bibinfo{pages}{1029} (\bibinfo{year}{1988}).

\bibitem[{\citenamefont{Mishra et~al.}(2002)\citenamefont{Mishra, Ma, and
  Zhang}}]{fczhang2002}
\bibinfo{author}{\bibfnamefont{A.}~\bibnamefont{Mishra}},
  \bibinfo{author}{\bibfnamefont{M.}~\bibnamefont{Ma}}, \bibnamefont{and}
  \bibinfo{author}{\bibfnamefont{F.-C.} \bibnamefont{Zhang}},
  \bibinfo{journal}{Phys. Rev. B} \textbf{\bibinfo{volume}{65}},
  \bibinfo{pages}{214411} (\bibinfo{year}{2002}).

\bibitem[{\citenamefont{Bossche et~al.}(2000)\citenamefont{Bossche, Zhang, and
  Mila}}]{fczhang2000}
\bibinfo{author}{\bibfnamefont{M.~V.~D.} \bibnamefont{Bossche}},
  \bibinfo{author}{\bibfnamefont{F.~C.} \bibnamefont{Zhang}}, \bibnamefont{and}
  \bibinfo{author}{\bibfnamefont{F.}~\bibnamefont{Mila}},
  \bibinfo{journal}{Euro. Phys. J. B} \textbf{\bibinfo{volume}{17}},
  \bibinfo{pages}{367} (\bibinfo{year}{2000}).

\bibitem[{\citenamefont{Chen et~al.}(2005)\citenamefont{Chen, Wu, Zhang, and
  Wang}}]{wu2005}
\bibinfo{author}{\bibfnamefont{S.}~\bibnamefont{Chen}},
  \bibinfo{author}{\bibfnamefont{C.}~\bibnamefont{Wu}},
  \bibinfo{author}{\bibfnamefont{S.-C.} \bibnamefont{Zhang}}, \bibnamefont{and}
  \bibinfo{author}{\bibfnamefont{Y.}~\bibnamefont{Wang}},
  \bibinfo{journal}{Phys. Rev. B} \textbf{\bibinfo{volume}{72}},
  \bibinfo{pages}{214428} (\bibinfo{year}{2005}).

\bibitem[{\citenamefont{Pankov et~al.}(2007)\citenamefont{Pankov, Moessner, and
  Sondhi}}]{pankov2007}
\bibinfo{author}{\bibfnamefont{S.}~\bibnamefont{Pankov}},
  \bibinfo{author}{\bibfnamefont{R.}~\bibnamefont{Moessner}}, \bibnamefont{and}
  \bibinfo{author}{\bibfnamefont{S.~L.} \bibnamefont{Sondhi}},
  \bibinfo{journal}{Phys. Rev. B} \textbf{\bibinfo{volume}{76}},
  \bibinfo{pages}{104436} (\bibinfo{year}{2007}).

\bibitem[{\citenamefont{Xu and Wu}(2008)}]{xuwu2008}
\bibinfo{author}{\bibfnamefont{C.}~\bibnamefont{Xu}} \bibnamefont{and}
  \bibinfo{author}{\bibfnamefont{C.}~\bibnamefont{Wu}}, \bibinfo{journal}{Phys.
  Rev. B} \textbf{\bibinfo{volume}{77}}, \bibinfo{pages}{134449}
  (\bibinfo{year}{2008}).

\bibitem[{\citenamefont{Wu}(2008)}]{congjun}
\bibinfo{author}{\bibfnamefont{C.}~\bibnamefont{Wu}}, \bibinfo{journal}{private
  communication}  (\bibinfo{year}{2008}).

\bibitem[{\citenamefont{Fradkin and Kivelson}(1990)}]{fradkin1990}
\bibinfo{author}{\bibfnamefont{E.}~\bibnamefont{Fradkin}} \bibnamefont{and}
  \bibinfo{author}{\bibfnamefont{S.~A.} \bibnamefont{Kivelson}},
  \bibinfo{journal}{Mod. Phys. Lett. B} \textbf{\bibinfo{volume}{4}},
  \bibinfo{pages}{225} (\bibinfo{year}{1990}).

\bibitem[{\citenamefont{Sachdev and Read}(1991)}]{sachdev1991}
\bibinfo{author}{\bibfnamefont{S.}~\bibnamefont{Sachdev}} \bibnamefont{and}
  \bibinfo{author}{\bibfnamefont{N.}~\bibnamefont{Read}},
  \bibinfo{journal}{Int. J. Mod. Phys. B} \textbf{\bibinfo{volume}{5}},
  \bibinfo{pages}{219} (\bibinfo{year}{1991}).

\bibitem[{\citenamefont{Lou et~al.}(2007)\citenamefont{Lou, Sandvik, and
  Balents}}]{sandvik2007a}
\bibinfo{author}{\bibfnamefont{J.}~\bibnamefont{Lou}},
  \bibinfo{author}{\bibfnamefont{A.~W.} \bibnamefont{Sandvik}},
  \bibnamefont{and} \bibinfo{author}{\bibfnamefont{L.}~\bibnamefont{Balents}},
  \bibinfo{journal}{Phys. Rev. Lett} \textbf{\bibinfo{volume}{99}},
  \bibinfo{pages}{207203} (\bibinfo{year}{2007}).

\bibitem[{\citenamefont{Motrunich and Vishwanath}(2008)}]{ashvin2008}
\bibinfo{author}{\bibfnamefont{O.~I.} \bibnamefont{Motrunich}}
  \bibnamefont{and}
  \bibinfo{author}{\bibfnamefont{A.}~\bibnamefont{Vishwanath}},
  \bibinfo{journal}{arXiv:0805.1494}  (\bibinfo{year}{2008}).

\end{thebibliography}
\end{document}